\documentclass[twocolumn,prl,tightenlines,superscriptaddress]{revtex4-2}
\usepackage{hyperref}
\usepackage{amsmath,amssymb}
\usepackage[utf8]{inputenc} 			
\usepackage[T1]{fontenc}				
\usepackage{graphicx}
\usepackage{float}
\usepackage[usenames,dvipsnames]{xcolor}
\usepackage{tikz}
\usepackage{ulem}
\usepackage{siunitx}

\renewcommand{\div}{\nabla\cdot}
\renewcommand{\vec}[1]{\mathbf{#1}}

\newcommand*\colvec[3][]{
    \begin{pmatrix}\ifx\relax#1\relax\else#1\\\fi#2\\#3\end{pmatrix}
}

\begin{document}

\title{Inescapable anisotropy of non-reciprocal XY models}

\author{Dawid Dopierala}
\affiliation{TCM Group, Cavendish Laboratory, JJ Thomson Avenue, Cambridge, CB3 0HE, UK}
\affiliation{Sorbonne Universit\'e, CNRS, Laboratoire de Physique Th\'eorique de la Mati\`ere Condens\'ee, 75005 Paris, France}
\author{Hugues Chat\'{e}}
\affiliation{Service de Physique de l'Etat Condens\'e, CEA, CNRS Universit\'e Paris-Saclay, CEA-Saclay, 91191 Gif-sur-Yvette, France}
\affiliation{Computational Science Research Center, Beijing 100094, China}
\affiliation{Sorbonne Universit\'e, CNRS, Laboratoire de Physique Th\'eorique de la Mati\`ere Condens\'ee, 75005 Paris, France}
\author{Xia-qing Shi}
\affiliation{Center for Soft Condensed Matter Physics and Interdisciplinary Research, Soochow University, Suzhou 215006, China}
\author{Alexandre Solon}
\affiliation{Sorbonne Universit\'e, CNRS, Laboratoire de Physique Th\'eorique de la Mati\`ere Condens\'ee, 75005 Paris, France}

\date{\today}
\begin{abstract}
  We investigate non-reciprocal XY (NRXY) models defined on two-dimensional lattices
  in which the coupling strength of a spin with its neighbors varies with their position 
  in the frame defined by the current spin orientation.
  As expected from the seminal work of Dadhichi et al., Phys. Rev. E 101, 052601
(2020), we first show that non-reciprocity is akin
  to a self-propulsion: we derive a mean-field continuous theory identical to that of constant-density flocks. 
  Like the latter, NRXY models exhibit a long-range ordered phase that is
  metastable to the nucleation of topological defects, and their
  asymptotic state is a dynamic foam of asters.
  We then show that in the metastable ordered phase, the lattice always induces anisotropy on large scales, 
  pinning the direction of order and imposing a finite correlation length.
  Crucially, we demonstrate that this anisotropy is inescapable since, even when not explicitly present in the model,
 it is generated by fluctuations. In short, the ordered phase of lattice NRXY models is that of active clock models.
\end{abstract}

\maketitle

Non-reciprocal interactions are ubiquitous out of equilibrium. They
have come under intense scrutiny because of the phenomenology they
generate.
In particular, mixing two populations with antagonistic goals can
result in a spectacular array of intriguing emergent phenomena and
dynamical
patterns~\cite{stenhammar_activity-induced_2015,wysocki_propagating_2016,saha_scalar_2020,you_nonreciprocity_2020,fruchart_non-reciprocal_2021,dinelli_non-reciprocity_2023,chatterjee_flocking_2023,duan_dynamical_2023,pisegna_emergent_2024,zelle_universal_2024,golestanian_non-reciprocal_2024,avni_dynamical_2024,duan_phase_2025}.
Non-reciprocity can also be present within a single population, such
as active systems featuring self-propelled particles with a front-back
asymmetry~\cite{dadhichi_nonmutual_2020}. This again leads to a
variety of striking collective effects, from milling and moving
aggregates when interactions mimick a field of
vision~\cite{pearce_role_2014,barberis_large-scale_2016,chen_fore-aft_2017},
to spontaneous breaking of chiral
symmetry~\cite{romanczuk_emergent_2016,fruchart_odd_2023}.

The non-reciprocal XY model (NRXY) is a simple system where such
single-species non-reciprocal interactions are at play. It has
attracted much attention recently, under the form of extensions of the
classic polar XY model on two-dimensional lattices where continuous
spins align more strongly with neighbors which their current
orientation points
to~\cite{dadhichi_nonmutual_2020,loos_long-range_2023,bandini_xy_2024,liu_dynamic_2025,rouzaire_non-reciprocal_2024}.
Even though they all find qualitative differences with the equilibrium
case, these works globally leave a confusing picture, particularly
about the nature of the low-temperature ordered phase.  For an
implementation with vision cones~\cite{loos_long-range_2023}, it was
first found to have long-ranged order (LRO) ---in contrast with the
quasi-long-range order (QLRO) with temperature-dependent exponents
present in equilibrium---, except for special opening angles of the
vision cone.  Ref.~\cite{loos_long-range_2023} also claimed an
additional QLRO region at intermediate temperatures on the hexagonal
lattice but not on the square lattice.  In contrast,
Ref.~\cite{bandini_xy_2024} found LRO rather than QLRO at the special
angles, while Ref.~\cite{liu_dynamic_2025} argued that a small QLRO
region is present on the square lattice for large opening angles
$\lesssim \ang{360}$.

In this Letter, we clarify the nature of the low-temperature phase in
NRXY models, and conclude that it does not depend on the lattice considered nor on special angles of vision.
In agreement with the seminal work of Dadhichi et al.~\cite{dadhichi_nonmutual_2020},
we first show that non-reciprocity amounts to a form of self-propulsion: we derive
a mean-field continuum description identical to that of constant density
flocks~\cite{toner_birth_2012,besse_metastability_2022}.
Consistently with our knowledge of such a theory, 
we then find that the low temperature homogeneous ordered phase of NRXY models 
debated in \cite{loos_long-range_2023,bandini_xy_2024,liu_dynamic_2025}
is actually metastable to the nucleation of defects so that, in
large-enough systems, the asymptotic phase is a dynamic foam of asters
akin to that reported in~\cite{besse_metastability_2022}.  The
metastable homogeneous ordered phase is found to always display LRO,
but not the non-trivial scale free correlations of constant-density
flocks~\cite{toner_birth_2012,chate_dynamic_2024}, which can only be
observed at intermediate scales at best.  The asymptotic LRO, because
of the lattice spatial anisotropy effects, is similar to that of
active clock
models~\cite{chatterjee_polar_2022,solon_susceptibility_2022},
exhibiting a finite correlation length. Interestingly, we find that
anisotropy is inescapable, since it is generated by fluctuations in
cases where it does not appear explicitly.

{\it NRXY models as constant density flocks.} 
The spin at lattice site $i$, located at position $\vec r_i$, 
carries an angle $\theta_i$ that interacts with neighboring sites $j$ and evolves under the Langevin dynamics: 
\begin{equation}
  \label{eq:model}
  \partial_t\theta_i=\sum_{jvi} J(\psi_{ij})\sin(\theta_j-\theta_i)+\sqrt{2T}\eta_i
\end{equation}
where $\eta_i$ is a unit variance Gaussian white noise, and
$J(\psi_{ij})$ is a coupling strength depending on
$\psi_{ij}=\theta_i-\arg(\vec r_j-\vec r_i)$, which can be seen as the
angular position of $j$ in the 'field of view' of $i$.~\footnote{The
  equilibrium XY model at temperature $T$ is recovered when $J$ is
  constant.}

We first derive the mean-field continuum description of the above NRXY
model.  Starting from Eq.~(\ref{eq:model}) and using the It\=o
formula, one gets an infinite hierarchy of evolution equations for the
angular harmonics $\langle \cos(p\theta_i)\rangle$ and
$\langle \sin(p\theta_i)\rangle$ where $p$ is an integer and the
average is taken over realizations of the noise.  We then use the
mean-field approximation
$\langle f(\theta_i)g(\theta_j)\rangle\approx\langle
f(\theta_i)\rangle\langle g(\theta_j)\rangle$ for $i\neq j$ and any
functions $f$ and $g$. Given the symmetries of the system, the order
parameter is expected to be the first harmonics, {\it i.e.} the
magnetization
$m_i=(\langle\cos\theta_i\rangle,\langle \sin\theta_i\rangle)$. We
thus use a standard closure scheme for polar
systems~\cite{peshkov_boltzmann-ginzburg-landau_2014,solon_susceptibility_2022},
neglecting harmonics with $p>2$ and enslaving the $p=2$ harmonics to
$m_i$. Finally, taking the continuum limit with the lattice spacing as
the unit length, we arrive at a closed equation for the magnetization
field $\vec m(\vec r)$
\begin{multline}
  \label{eq:MF}
  \partial_t \vec m+\lambda_1 \vec m\cdot\nabla\vec m+\lambda_2 (\div \vec m)\vec m\\+\lambda_3 \nabla(|\vec m|^2)=D\nabla^2 \vec m+(\alpha-\gamma |\vec m|^2)\vec m
\end{multline}
where the parameters depend on the lattice considered and on the coefficients of the Fourier expansion 
$J(\psi)=J_0+\sum_{p=1}^{+\infty}J_p \cos(p\psi)$.~\footnote{Above we omitted a term 
$\sim J_2 (\partial_x^2-\partial_y^2)(-m_x,m_y)$ which has negligible effects and vanishes for the coupling function without
nematic harmonics.}
(Below we mostly use a square latttice, for which
$\lambda_1=\frac{J_1}{2}\left(-3+\frac{J_0}{T}\right)$,
$\lambda_2=\frac{J_1}{2}\left(1-\frac{J_0}{T}\right)$,
$\lambda_3=\frac{J_1}{4}\left(1+\frac{J_0}{T}\right)$,
$\alpha=2J_0-T$, $\gamma=\frac{2 J_0^2}{T}$ and
$D=\frac{J_0}{2}$.)

Eq.~(\ref{eq:MF}) is a constant-density version of the Toner-Tu
equations~\cite{toner_long-range_1995} describing polar flocks.  Its
emergence here is a clear realization of the idea put forward in
Ref.~\cite{dadhichi_nonmutual_2020} that non-reciprocal interactions
act in a similar way as self-propulsion.  Eq.\eqref{eq:MF} equipped
with a noise term has an homogeneous flocking phase with non-trivial
long-range correlations
\cite{toner_birth_2012,chate_dynamic_2024}. This phase, however, is
metastable to the nucleation of defects, so that the asymptotic regime
of~\eqref{eq:MF} in the low-noise regime is a dynamic foam of
asters~\cite{besse_metastability_2022}.

We now show that most of this phenomenology is found in lattice NRXY
models.  In the examples shown below we consider a square lattice,
restrict the interaction to four nearest neighbors, and mostly use two
different $J(\psi_{ij})$ kernels with stronger coupling in front of
the spins: a minimal one that includes only the first harmonics
\begin{equation}
  \label{eq:J1}
J^{(1)}(\psi_{ij})=J_0+J_1\cos(\psi_{ij}),
\end{equation}
and the vision cones of opening angle $\Psi$ of \cite{loos_long-range_2023,bandini_xy_2024,liu_dynamic_2025}
\begin{equation}
  \label{eq:JVC}
J^{\rm VC}(\psi_{ij})=J_0,\,\text{if}\,\min(|\psi_{ij}|,2\pi-|\psi_{ij}|)\le \Psi/2
\end{equation}
and $0$ otherwise.  We consider only ferromagnetic alignment $J_0>0$
and $J_1\le J_0$ and $J_1>0$ without loss of generality.  The ``smooth
vision cones'' of Ref.~\cite{rouzaire_non-reciprocal_2024} can be
regarded as intermediate between Eqs.~(\ref{eq:J1})
and~(\ref{eq:JVC}).

\begin{figure}
  \centering
  \includegraphics[width=\columnwidth]{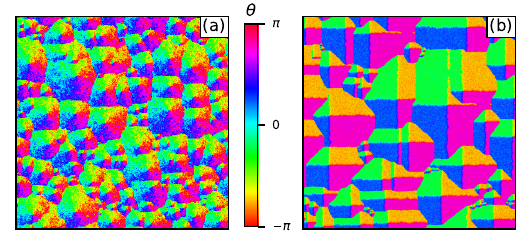}\\
  \includegraphics[width=0.49\columnwidth]{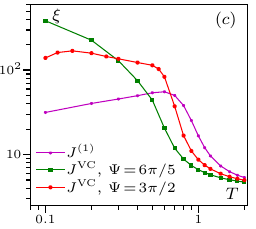}
  \includegraphics[width=0.49\columnwidth]{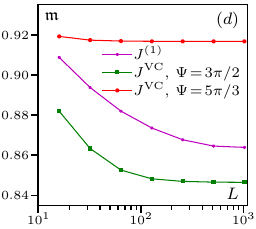}
  \caption{(a,b) Snapshots in the active foam phase with the $J^{(1)}$
    kernel (a) and vision cones $J^{\rm VC}$ with $\Psi=6\pi/5$ (b). 
    (c) Correlation length as a function of temperature in
    the foam phase. (b): Order parameter vs system
    size in the homogeneous ordered phase. 
    Parameters: $L=400$ (a), $L=2048$ (b,c), $T=0.3$ (a), $T=0.1$ (b), $T=0.3$ (d). For all
    panels: $J_0=1$, $J_1=1$ for the $J^{(1)}$ kernel and $dt=0.1$}
  \label{fig:cst-density-flocks}
\end{figure}

Starting from a disordered initial condition, the
stationary state is always, after transient, and in a large-enough system, 
a dynamic foam of domains with inward pointing radial order around
aster defect cores.  These domains are delimited by shock lines at the
intersection of which new defects are constantly nucleated
(Fig.~\ref{fig:cst-density-flocks}(a,b), Movies 1 \& 2 in
\cite{SUPP}). For kernel $J^{(1)}$, asters are nearly-isotropic and
shocklines are found in all directions, as for constant-density
flocks, but for $J^{VC}$ the patterns are clearly anisotropic, with
most spins pointing along the diagonals and shocklines oriented
preferentially along the lattice axes.

The correlation length $\xi$~\footnote{Defined as
  $\xi=\int_0^\pi S(q)dq/\int_0^\pi qS(q)dq$ with the structure factor
  $S(q)=\langle \vec m(\vec q)\vec m(-q)\rangle$ averaged over time
  and orientations of $\vec q$.} provides an estimate of the typical
defect size (Fig.~\ref{fig:cst-density-flocks}(c)). As for
constant-density flocks~\cite{besse_metastability_2022}, it does not
show any sign of divergence at a finite temperature that would signal
a binding/unbinding transition for the topological defects.
Note that we were able to observe a steady state foam of asters down
to $T=0.1$, much lower than the order-disorder transition temperature
reported for small systems~\cite{loos_long-range_2023}.

As expected, the homogeneous ordered phase debated in
\cite{loos_long-range_2023,bandini_xy_2024,liu_dynamic_2025} is
metastable to the nucleation of an aster and its accompanying shock
line, an event which triggers a cascade of further nucleations leading
eventually to the steady-state dynamic foam described above (Movies 3
\& 4 in \cite{SUPP}).  We thus conclude that it seems to be the only
asymptotic low-temperature phase of NRXY models.

The long-ranged ordered metastable phase can nevertheless be observed and studied
in large systems, provided that one discards data following the first successful nucleation event.
Computing the mean magnetization
per spin $\mathfrak{m}=\langle |\sum_i\vec u(\theta_i)|\rangle/N$ with
$\vec u(\theta_i)=(\cos\theta_i,\sin\theta_i)$, we see that it reaches a
non-zero value at large system sizes (see
Fig.~\ref{fig:cst-density-flocks}(d)), signaling long-range
order. This stands even for vision cones with opening angles
$\Psi=n\pi/2$ for which QLRO was reported in Ref.~\cite{loos_long-range_2023} based on
simulations of smaller systems.

{\it Explicit anisotropy.}  Compared to the minimal kernel $J^{(1)}$,
vision cones such as those used in
Fig.~\ref{fig:cst-density-flocks}(b) have higher-order harmonics
$\cos(p\psi_{ij})$ with $p>1$. In particular, when $p$ is a multiple
of $4$, for a square lattice $\cos(p\psi_{ij})=\cos(p\theta_i)$
becomes an explicit anisotropic coupling strength independent of spin
$j$. For clarity, let us isolate this ingredient by considering a
minimally anisotropic kernel that contains only the $4$-th harmonic
\begin{equation}
  \label{eq:J4}
J^{(4)}(\psi_{ij})=J_0+J_4\cos\left(4\psi_{ij}\right)=J_0+J_4\cos\left(4\theta_i\right).
\end{equation}
with $|J_4|<J_0$. Concretely, with this interaction kernel, spin $i$
has a larger (resp. smaller) coupling strength with all its neighbors
when it points along a direction of the lattice when $J_4>0$ (resp
$J_4<0$).

This leads to the pinning of the direction of order
along favored directions. To show this, we consider the mean direction
$\bar\theta=\frac{1}{N}\sum_i\theta_i$ 
which, using Eq.~(\ref{eq:model}) and kernel~(\ref{eq:J4}) evolves exactly as
\begin{equation}
  \label{eq:bartheta}
\partial_t\bar\theta=\frac{J_4}{N}\sum_{ivj} \left[\cos(4\theta_i)\!\!-\!\!\cos(4\theta_j)\right]\sin(\theta_j\!\!-\!\!\theta_i)+\sqrt{\frac{2T}{N}}\eta
\end{equation}
where the sum is over all bonds and $\eta$ is a unit variance Gaussian
white noise.  Taking the continuum limit, we then expand the
trigonometric functions around $\phi\equiv\langle\bar\theta\rangle$
(brackets denote noise average), writing
$\theta(\vec r,t)=\phi(t)+\delta\theta(\vec r,t)$, assuming we are at
low temperature so that deviations are small. In the spirit of
Refs.~\cite{martin_fluctuation-induced_2021,martin_fluctuation-induced_2024},
we average over the fluctuations to obtain, at leading order, the
following deterministic evolution for the mean direction
\begin{equation}
  \label{eq:bartheta-av}
\partial_t\phi=4 J_4\sin(4\phi)\langle\frac{1}{L^2}\int d\vec x |\nabla\delta\theta|^2\rangle.
\end{equation}
The integral on the r.h.s. of Eq.~(\ref{eq:bartheta-av}) can be
evaluated by linearizing 
Eq.~(\ref{eq:model}) which yields in the continuum
\begin{equation}
  \label{eq:linear-cont}
  \partial_t\delta\theta(\vec r,t)=(J_0+J_4\cos(4\phi))\Delta\delta\theta+\sqrt{2T}\eta(\vec r,t)
\end{equation}
with $\eta(\vec r,t)$ a unit variance Gaussian white noise field.  On
the fast timescale of fluctuations, $\phi$ is constant and
Eq.~(\ref{eq:linear-cont}) is an equilibrium dynamics, so the value of
the integral in Eq.~(\ref{eq:bartheta-av}) is directly given by the
equipartition of energy
$\langle\frac{1}{L^2}\int d\vec x
|\nabla\delta\theta|^2\rangle=T/(J_0+J_4\cos(4\phi))$. We thus find
that the global direction evolves as
\begin{equation}
  \label{eq:bartheta-av-pot}
  \partial_t\phi=\frac{4 J_4T\sin(4\phi)}{J_0+J_4\cos(4\phi)}=-\frac{d V}{d\phi}
\end{equation}
where the last equality defines the potential $V(\phi)$.

\begin{figure}
  \centering
  \includegraphics[width=\columnwidth]{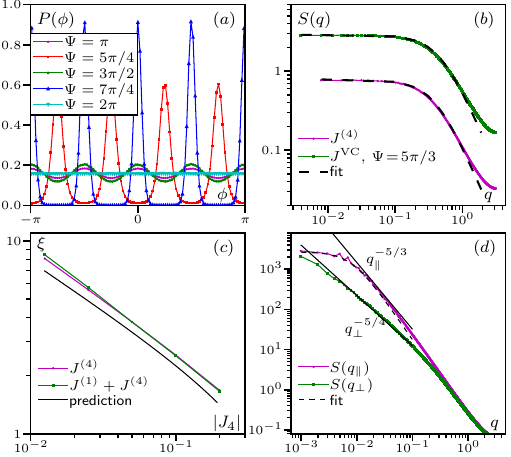}
  \caption{(a) Histogram of the global direction of order $\phi$ in a
    system small enough ($L=10$) that it explores all directions
    (visions cones with various angles $\Psi$).  (b) Structure factor
    $S(q)=\langle |\theta(\vec q)|^2\rangle$ radially averaged fitted
    by function $c/(1+(q\xi)^2)$ used to extract the correlation
    length $\xi$.  (c): Correlation length vs anisotropy coefficient
    $J_4$ for the $J^{(4)}$ kernel Eq.~(\ref{eq:J4}) and the same
    kernel complemented with the $J_1$ term of Eq.~(\ref{eq:J1}),
    compared to the prediction Eq.~(\ref{eq:xi}).  (d): Structure
    factor in the longitudinal and transverse directions for the
    $J^{(1)}$ kernel. Solid black lines indicate the scaling exponents
    for constant density flocks~\cite{chate_dynamic_2024}. The dashed
    line is the same fit as in panel (b) used to extract
    $\xi$. Parameters: (a): $T=0.35$, $L=10$. (b): $T=0.1$ and $L=800$
    (magenta), $T=0.4$, $L=1600$ (green). (c): $T=0.1$,
    $L=800$. (d): $T=0.2$, $L=6400$. $dt=0.1$, $J_0=J_1=1$
    everywhere.}
  \label{fig:anisotropy}
\end{figure}

For positive $J_4$, the directions of the lattice
$\phi=n\pi/2$ are maxima of the potential $V(\phi)$ while the minima,
where the system is expected to settle, correspond to the
diagonals. We thus obtain the apparently counter-intuitive result that
the preferred directions are those where alignment is the
weakest. Nevertheless, this can be rationalized easily: if the mean
direction $\phi$ is such that the spins have a minimal coupling
strength then a spin which fluctuates away from $\phi$ will have a
larger coupling and thus will align stronger with $\phi$ than the rest
of the system aligns with it. Hence, there is a net force bringing
back the spin to the direction of weakest alignment, on which the
order is thus pinned. This is fully consistent with
numerical results. In particular, for vision cones, one can extract
the fourth harmonics of the kernel
\begin{equation}
  \label{eq:4th-harmo}
 J_4^{\rm VC}=\frac{1}{2\pi}\int J^{\rm  VC}(\theta)\cos(4\theta)d\theta=\frac{\sin(2\Psi)}{4\pi}.
\end{equation}
Consequently the order is pinned on the directions of the lattice for
opening angles such that $\sin(2\Psi)<0$ and in the diagonals when
$\sin(2\Psi)>0$ as shown in Fig.~\ref{fig:anisotropy}
(a)~\footnote{Interestingly, this depends on the type of dynamics: For
  Monte-Carlo dynamics in which the angle can make arbitrary jumps at
  each time step, the pinning directions are
  opposite~\cite{loos_long-range_2023,bandini_xy_2024} compared to our
  Langevin dynamics Eq.~(\ref{eq:model}).}.

Importantly. contrary to constant density flocks which are scale
invariant in the ordered phase, the anisotropy exhibited above induces
a finite correlation length to NRXY models, as seen numerically on the
correlation functions (Fig.~\ref{fig:anisotropy}(b)).  To estimate the
correlation length analytically, let us consider the fluctuations
around a stable direction (we choose $J_4<0$ and $\phi=0$) and make
the mean-field-like approximation that the direction $\theta$ is
pinned by the same potential $V(\theta)$ defined in
Eq.~(\ref{eq:bartheta-av-pot}). At linear order one then gets
\begin{equation}
  \label{eq:linear-cont-zero}
  \partial_t\delta\theta(\vec r,t)=(J_0+J_4)\Delta\delta\theta-V''(0)\delta\theta+\sqrt{2T}\eta(\vec r,t)
\end{equation}
which gives a correlation length 
\begin{equation}
  \label{eq:xi}
 \xi=\sqrt{\frac{J_0+J_4}{V''(0)}}=\frac{J_0+J_4}{\sqrt{16|J_4|T}}.
\end{equation}
As seen in Fig.~\ref{fig:anisotropy} (c), Eq.~(\ref{eq:xi}) is within
a factor $1.2$ of the measured value and thus provides a relatively
good estimate considering the approximations made.

{\it Fluctuation-induced anisotropy.}  We have shown that explicitly
anisotropic coupling kernels lead to pinning the direction of
order and long-range order with a finite correlation length. 
For the models without such an anisotropy, such as the kernel
$J^{(1)}$ of Eq.~(\ref{eq:J1}) and the vision cones with $\psi=n\pi/2$
(for which Eq.~(\ref{eq:4th-harmo}) gives $J_4=0$), one does not observe 
rotational symmetry with no preferred direction of ordering: 
as shown in Fig.~\ref{fig:anisotropy}(a),
although the pinning is much weaker, one observes that vision cones
with $\psi=n\pi/2$ still favor ordering along the directions of the
lattice. For the $J^{(1)}$ kernel, global order is also
pinned (not shown) and the structure factor saturates at small
wavevectors (Fig.~\ref{fig:anisotropy} (d)).

Here, the observed lattice effects result from a subtle interplay
between correlations, which are well known to be anisotropic with
respect to the direction of
order~\cite{toner_birth_2012,chate_dynamic_2024}, and the breaking of
rotational invariance by the lattice structure. Let us see how this
works in practice by following the  approach that led to
Eq.~(\ref{eq:linear-cont-zero}). Writing the dynamics of $\bar\theta$
as in Eq.~(\ref{eq:bartheta}) but for the $J^{(1)}$ kernel gives the
exact equation
\begin{multline}
  \label{eq:bartheta-J1}
  \partial_t\bar\theta=\frac{J_1}{N}\sum_{i,j} \left[\left(\cos(\theta_{i,j})\!+\!\cos(\theta_{i+1,j})\right)\sin(\theta_{i+1,j}\!-\!\theta_{ij})\right.\\ \left.+\left(\cos(\theta_{i,j})\!+\!\cos(\theta_{i,j+1})\right)\sin(\theta_{i,j+1}\!-\!\theta_{ij})\right]\!+\!\sqrt{\frac{2T}{N}}\eta
\end{multline}
where now $i$ and $j$ are indices on the lattice and the sum is over
all lattice sites. As before, we then take the continuum limit, expand
around the average orientation
$\theta(\vec r,t)=\phi(t)+\delta\theta(\vec r,t)$ with
$\phi=\langle\bar\theta\rangle$ and average over the fluctuations. To
lighten the discussion, let us further assume without loss of
generality that $\phi\approx 0$ so that the correlation of
fluctuations can be evaluated at $\phi=0$ to leading order. Then by
the symmetry $\theta\to-\theta, y\to -y$, all correlations that
contain an odd number of $\theta$ and $y$-derivatives vanish.
Moreover all terms that can be written as total derivatives integrate
to $0$, so that at leading order there remains
\begin{equation}
  \label{eq:phi-J1}
  \partial_t\phi=-r\phi; \quad r=J_1\left(\tfrac{1}{6}\langle\delta\theta^2\partial_y^3\delta\theta\rangle_0-\tfrac{1}{3}\langle(\partial_y\delta\theta)^3\rangle_0\right)
\end{equation}
with the subscript $0$ on the noise average indicating that it should
be computed at $\phi=0$.

The correlators in Eq.~(\ref{eq:phi-J1}) can be measured numerically
to check the consistency of the approach. For the parameters of
Fig.~\ref{fig:anisotropy}(d), we measured $r\approx 6\times 10^{-4}$
from which one predicts a correlation length
$\xi=\sqrt{J_0/r}\approx 40$ which is a factor $3$ smaller than the
value fitted in Fig.~\ref{fig:anisotropy} but still gives the right
order of magnitude. Note that the non-vanishing value of $r$ does come
from the fact that fluctuations are coupled to the direction. In the
equilibrium XY model, $r=0$ because of the symmetry
$\theta\to -\theta$.

\begin{figure}
  \centering
  \includegraphics[width=\columnwidth]{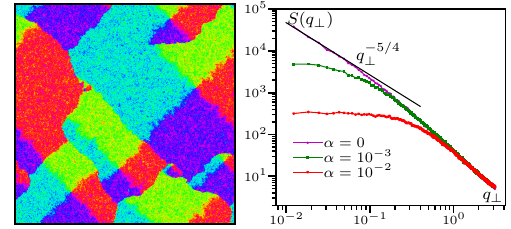}
  \caption{ Active clock continuum equations.  Left: Snapshot taken in
    the aster foam phase. Same color code as
    Fig.~\ref{fig:cst-density-flocks}(a,b).  Right: Structure factor
    in the ordered phase for wavevectors transverse to the direction
    of order.  The black line indicates the exponent computed
    in~\cite{chate_dynamic_2024} for constant density flocks.
    Parameters: L=1024, $J_0=J_1=1$, $T=1.6$ (left), $T=1.5$ (right),
    $\alpha=10^{-2}$ (left), $\eta=0.4$. Integrated using a
    semi-spectral algorithm with $3/2$ anti-aliasing ($dx=1$) and
    Euler time stepping ($dt=0.1$).  }
  \label{fig:clock}
\end{figure}

{\it Active clock model.} We have shown that NRXY models are
generically anisotropic, either explicitly or implicitly under the
effect of fluctuations. As shown in~\cite{solon_susceptibility_2022},
anisotropy is always relevant at large scales for active systems that
possess long-range order.  The behavior of NRXY models is thus
expected to be that of an active constant-density clock model, which
can be described, following \cite{solon_susceptibility_2022}, by
adding, on the r.h.s. of the continuum Eq.~(\ref{eq:MF}), a pinning
force $-\partial_\theta V(\theta) \vec m_\bot$ that acts in direction
$\vec m_\bot=(-m_y,m_x)$ transverse to the local magnetization with
$V(\theta)=-\alpha\cos(n\theta)$ (here $\theta=\arg(\vec m)$
and $n$ the number of lattice directions).

Simulating the above theory with a Gaussian white noise term
$\sqrt{2T}\eta(\vec r,t)$ indeed recapitulates the phenomenology of
NRXY models, in particular the dynamic foam of asters and a structure
factor equal to that of constant density flocks on short scales but
dominated by anisotropy on large scales in the ordered phase
(Fig.~\ref{fig:clock} and Movie 5 in \cite{SUPP}).

{\it Conclusion.}  NRXY models combine two ingredients, activity and
anisotropy.  (i) Activity: non-reciprocity, by coupling the direction
of the spins to spatial directions, has an effect similar to
self-propulsion. We have shown that NRXY models and constant density
flocks~\cite{toner_birth_2012,besse_metastability_2022} share the same
mean-field description, and display its main features such as a
low-temperature state with LRO and non-trivial scaling of correlations
that is metastable to the nucleation of topological defects. (ii)
Inescapable anisotropy due to the lattice: we demonstrated how, even
in models that are not explicitly anisotropic, anisotropy is generated
by fluctuations. Because it is always a relevant perturbation in the
renormalisation group sense~\cite{solon_susceptibility_2022}, it
dominates the large scale physics, however weak it may be.  Putting
these two ingredients together, NRXY models can be thought off as
constant density active clock models, of which we have proposed a
continuum theory.

This clear picture allows us to discuss some of the observations
reported in previous publications. First, although this is apparent
only beyond a possibly large crossover length, the defectless phase
always has LRO and never QLRO as reported for vision cones on
hexagonal lattice~\cite{loos_long-range_2023} and for $\Psi$
close to $2\pi$~\cite{liu_dynamic_2025}. On the contrary, for
equilibrium models such as the ARXY and
SRXY (asymmetric- and symmetric-reciprocal XY) models of ~\cite{bandini_xy_2024}, 
which use reciprocal vision cones, one can
have LRO or QLRO depending on both the temperature and symmetry of the
lattice~\cite{jose_renormalization_1977,lapilli_universality_2006}, just like for equilibrium clock models. 
This reflects the fundamental role of activity, here in the form of
non-reciprocity, that changes the nature of the low-temperature phase,

Lattice anisotropy, being always relevant, is an unpleasant feature
for whom would like to to study the scale-free ordered state of NRXY
models. Future research will consider different ways to reduce the
anisotropy, from using local interactions spanning larger ranges, to
using disordered hyperuniform substrates.

{\it Note.} A preprint with the same conclusions regarding the
anisotropy of NRXY models~\cite{popli_dont_2025} was released while finalizing
the present manuscript.

\acknowledgements We thank David Martin for interesting discussions.

\bibliography{biblio.bib}

\begin{thebibliography}{40}%
\makeatletter
\providecommand \@ifxundefined [1]{%
 \@ifx{#1\undefined}
}%
\providecommand \@ifnum [1]{%
 \ifnum #1\expandafter \@firstoftwo
 \else \expandafter \@secondoftwo
 \fi
}%
\providecommand \@ifx [1]{%
 \ifx #1\expandafter \@firstoftwo
 \else \expandafter \@secondoftwo
 \fi
}%
\providecommand \natexlab [1]{#1}%
\providecommand \enquote  [1]{``#1''}%
\providecommand \bibnamefont  [1]{#1}%
\providecommand \bibfnamefont [1]{#1}%
\providecommand \citenamefont [1]{#1}%
\providecommand \href@noop [0]{\@secondoftwo}%
\providecommand \href [0]{\begingroup \@sanitize@url \@href}%
\providecommand \@href[1]{\@@startlink{#1}\@@href}%
\providecommand \@@href[1]{\endgroup#1\@@endlink}%
\providecommand \@sanitize@url [0]{\catcode `\\12\catcode `\$12\catcode
  `\&12\catcode `\#12\catcode `\^12\catcode `\_12\catcode `\%12\relax}%
\providecommand \@@startlink[1]{}%
\providecommand \@@endlink[0]{}%
\providecommand \url  [0]{\begingroup\@sanitize@url \@url }%
\providecommand \@url [1]{\endgroup\@href {#1}{\urlprefix }}%
\providecommand \urlprefix  [0]{URL }%
\providecommand \Eprint [0]{\href }%
\providecommand \doibase [0]{https://doi.org/}%
\providecommand \selectlanguage [0]{\@gobble}%
\providecommand \bibinfo  [0]{\@secondoftwo}%
\providecommand \bibfield  [0]{\@secondoftwo}%
\providecommand \translation [1]{[#1]}%
\providecommand \BibitemOpen [0]{}%
\providecommand \bibitemStop [0]{}%
\providecommand \bibitemNoStop [0]{.\EOS\space}%
\providecommand \EOS [0]{\spacefactor3000\relax}%
\providecommand \BibitemShut  [1]{\csname bibitem#1\endcsname}%
\let\auto@bib@innerbib\@empty
\bibitem [{\citenamefont {Stenhammar}\ \emph {et~al.}(2015)\citenamefont
  {Stenhammar}, \citenamefont {Wittkowski}, \citenamefont {Marenduzzo},\ and\
  \citenamefont {Cates}}]{stenhammar_activity-induced_2015}%
  \BibitemOpen
  \bibfield  {author} {\bibinfo {author} {\bibfnamefont {J.}~\bibnamefont
  {Stenhammar}}, \bibinfo {author} {\bibfnamefont {R.}~\bibnamefont
  {Wittkowski}}, \bibinfo {author} {\bibfnamefont {D.}~\bibnamefont
  {Marenduzzo}},\ and\ \bibinfo {author} {\bibfnamefont {M.~E.}\ \bibnamefont
  {Cates}},\ }\bibfield  {title} {\bibinfo {title} {Activity-induced phase
  separation and self-assembly in mixtures of active and passive particles},\
  }\href@noop {} {\bibfield  {journal} {\bibinfo  {journal} {Physical review
  letters}\ }\textbf {\bibinfo {volume} {114}},\ \bibinfo {pages} {018301}
  (\bibinfo {year} {2015})}\BibitemShut {NoStop}%
\bibitem [{\citenamefont {Wysocki}\ \emph {et~al.}(2016)\citenamefont
  {Wysocki}, \citenamefont {Winkler},\ and\ \citenamefont
  {Gompper}}]{wysocki_propagating_2016}%
  \BibitemOpen
  \bibfield  {author} {\bibinfo {author} {\bibfnamefont {A.}~\bibnamefont
  {Wysocki}}, \bibinfo {author} {\bibfnamefont {R.~G.}\ \bibnamefont
  {Winkler}},\ and\ \bibinfo {author} {\bibfnamefont {G.}~\bibnamefont
  {Gompper}},\ }\bibfield  {title} {\bibinfo {title} {Propagating interfaces in
  mixtures of active and passive {Brownian} particles},\ }\href@noop {}
  {\bibfield  {journal} {\bibinfo  {journal} {New Journal of Physics}\ }\textbf
  {\bibinfo {volume} {18}},\ \bibinfo {pages} {123030} (\bibinfo {year}
  {2016})}\BibitemShut {NoStop}%
\bibitem [{\citenamefont {Saha}\ \emph {et~al.}(2020)\citenamefont {Saha},
  \citenamefont {Agudo-Canalejo},\ and\ \citenamefont
  {Golestanian}}]{saha_scalar_2020}%
  \BibitemOpen
  \bibfield  {author} {\bibinfo {author} {\bibfnamefont {S.}~\bibnamefont
  {Saha}}, \bibinfo {author} {\bibfnamefont {J.}~\bibnamefont
  {Agudo-Canalejo}},\ and\ \bibinfo {author} {\bibfnamefont {R.}~\bibnamefont
  {Golestanian}},\ }\bibfield  {title} {\bibinfo {title} {Scalar active
  mixtures: {The} nonreciprocal {Cahn}-{Hilliard} model},\ }\href@noop {}
  {\bibfield  {journal} {\bibinfo  {journal} {Physical Review X}\ }\textbf
  {\bibinfo {volume} {10}},\ \bibinfo {pages} {041009} (\bibinfo {year}
  {2020})},\ \bibinfo {note} {publisher: APS}\BibitemShut {NoStop}%
\bibitem [{\citenamefont {You}\ \emph {et~al.}(2020)\citenamefont {You},
  \citenamefont {Baskaran},\ and\ \citenamefont
  {Marchetti}}]{you_nonreciprocity_2020}%
  \BibitemOpen
  \bibfield  {author} {\bibinfo {author} {\bibfnamefont {Z.}~\bibnamefont
  {You}}, \bibinfo {author} {\bibfnamefont {A.}~\bibnamefont {Baskaran}},\ and\
  \bibinfo {author} {\bibfnamefont {M.~C.}\ \bibnamefont {Marchetti}},\
  }\bibfield  {title} {\bibinfo {title} {Nonreciprocity as a generic route to
  traveling states},\ }\href@noop {} {\bibfield  {journal} {\bibinfo  {journal}
  {Proceedings of the National Academy of Sciences}\ }\textbf {\bibinfo
  {volume} {117}},\ \bibinfo {pages} {19767} (\bibinfo {year} {2020})},\
  \bibinfo {note} {publisher: National Acad Sciences}\BibitemShut {NoStop}%
\bibitem [{\citenamefont {Fruchart}\ \emph {et~al.}(2021)\citenamefont
  {Fruchart}, \citenamefont {Hanai}, \citenamefont {Littlewood},\ and\
  \citenamefont {Vitelli}}]{fruchart_non-reciprocal_2021}%
  \BibitemOpen
  \bibfield  {author} {\bibinfo {author} {\bibfnamefont {M.}~\bibnamefont
  {Fruchart}}, \bibinfo {author} {\bibfnamefont {R.}~\bibnamefont {Hanai}},
  \bibinfo {author} {\bibfnamefont {P.~B.}\ \bibnamefont {Littlewood}},\ and\
  \bibinfo {author} {\bibfnamefont {V.}~\bibnamefont {Vitelli}},\ }\bibfield
  {title} {\bibinfo {title} {Non-reciprocal phase transitions},\ }\href@noop {}
  {\bibfield  {journal} {\bibinfo  {journal} {Nature}\ }\textbf {\bibinfo
  {volume} {592}},\ \bibinfo {pages} {363} (\bibinfo {year} {2021})},\ \bibinfo
  {note} {publisher: Nature Publishing Group}\BibitemShut {NoStop}%
\bibitem [{\citenamefont {Dinelli}\ \emph {et~al.}(2023)\citenamefont
  {Dinelli}, \citenamefont {O’Byrne}, \citenamefont {Curatolo}, \citenamefont
  {Zhao}, \citenamefont {Sollich},\ and\ \citenamefont
  {Tailleur}}]{dinelli_non-reciprocity_2023}%
  \BibitemOpen
  \bibfield  {author} {\bibinfo {author} {\bibfnamefont {A.}~\bibnamefont
  {Dinelli}}, \bibinfo {author} {\bibfnamefont {J.}~\bibnamefont {O’Byrne}},
  \bibinfo {author} {\bibfnamefont {A.}~\bibnamefont {Curatolo}}, \bibinfo
  {author} {\bibfnamefont {Y.}~\bibnamefont {Zhao}}, \bibinfo {author}
  {\bibfnamefont {P.}~\bibnamefont {Sollich}},\ and\ \bibinfo {author}
  {\bibfnamefont {J.}~\bibnamefont {Tailleur}},\ }\bibfield  {title} {\bibinfo
  {title} {Non-reciprocity across scales in active mixtures},\ }\href@noop {}
  {\bibfield  {journal} {\bibinfo  {journal} {Nature Communications}\ }\textbf
  {\bibinfo {volume} {14}},\ \bibinfo {pages} {7035} (\bibinfo {year}
  {2023})},\ \bibinfo {note} {publisher: Nature Publishing Group UK
  London}\BibitemShut {NoStop}%
\bibitem [{\citenamefont {Chatterjee}\ \emph {et~al.}(2023)\citenamefont
  {Chatterjee}, \citenamefont {Mangeat}, \citenamefont {Woo}, \citenamefont
  {Rieger},\ and\ \citenamefont {Noh}}]{chatterjee_flocking_2023}%
  \BibitemOpen
  \bibfield  {author} {\bibinfo {author} {\bibfnamefont {S.}~\bibnamefont
  {Chatterjee}}, \bibinfo {author} {\bibfnamefont {M.}~\bibnamefont {Mangeat}},
  \bibinfo {author} {\bibfnamefont {C.-U.}\ \bibnamefont {Woo}}, \bibinfo
  {author} {\bibfnamefont {H.}~\bibnamefont {Rieger}},\ and\ \bibinfo {author}
  {\bibfnamefont {J.~D.}\ \bibnamefont {Noh}},\ }\bibfield  {title} {\bibinfo
  {title} {Flocking of two unfriendly species: {The} two-species {Vicsek}
  model},\ }\href@noop {} {\bibfield  {journal} {\bibinfo  {journal} {Physical
  Review E}\ }\textbf {\bibinfo {volume} {107}},\ \bibinfo {pages} {024607}
  (\bibinfo {year} {2023})},\ \bibinfo {note} {publisher: APS}\BibitemShut
  {NoStop}%
\bibitem [{\citenamefont {Duan}\ \emph {et~al.}(2023)\citenamefont {Duan},
  \citenamefont {Agudo-Canalejo}, \citenamefont {Golestanian},\ and\
  \citenamefont {Mahault}}]{duan_dynamical_2023}%
  \BibitemOpen
  \bibfield  {author} {\bibinfo {author} {\bibfnamefont {Y.}~\bibnamefont
  {Duan}}, \bibinfo {author} {\bibfnamefont {J.}~\bibnamefont
  {Agudo-Canalejo}}, \bibinfo {author} {\bibfnamefont {R.}~\bibnamefont
  {Golestanian}},\ and\ \bibinfo {author} {\bibfnamefont {B.}~\bibnamefont
  {Mahault}},\ }\bibfield  {title} {\bibinfo {title} {Dynamical pattern
  formation without self-attraction in quorum-sensing active matter: {The}
  interplay between nonreciprocity and motility},\ }\href@noop {} {\bibfield
  {journal} {\bibinfo  {journal} {Physical Review Letters}\ }\textbf {\bibinfo
  {volume} {131}},\ \bibinfo {pages} {148301} (\bibinfo {year} {2023})},\
  \bibinfo {note} {publisher: APS}\BibitemShut {NoStop}%
\bibitem [{\citenamefont {Pisegna}\ \emph {et~al.}(2024)\citenamefont
  {Pisegna}, \citenamefont {Saha},\ and\ \citenamefont
  {Golestanian}}]{pisegna_emergent_2024}%
  \BibitemOpen
  \bibfield  {author} {\bibinfo {author} {\bibfnamefont {G.}~\bibnamefont
  {Pisegna}}, \bibinfo {author} {\bibfnamefont {S.}~\bibnamefont {Saha}},\ and\
  \bibinfo {author} {\bibfnamefont {R.}~\bibnamefont {Golestanian}},\
  }\bibfield  {title} {\bibinfo {title} {Emergent polar order in nonpolar
  mixtures with nonreciprocal interactions},\ }\href@noop {} {\bibfield
  {journal} {\bibinfo  {journal} {Proceedings of the National Academy of
  Sciences}\ }\textbf {\bibinfo {volume} {121}},\ \bibinfo {pages}
  {e2407705121} (\bibinfo {year} {2024})},\ \bibinfo {note} {publisher:
  National Academy of Sciences}\BibitemShut {NoStop}%
\bibitem [{\citenamefont {Zelle}\ \emph {et~al.}(2024)\citenamefont {Zelle},
  \citenamefont {Daviet}, \citenamefont {Rosch},\ and\ \citenamefont
  {Diehl}}]{zelle_universal_2024}%
  \BibitemOpen
  \bibfield  {author} {\bibinfo {author} {\bibfnamefont {C.~P.}\ \bibnamefont
  {Zelle}}, \bibinfo {author} {\bibfnamefont {R.}~\bibnamefont {Daviet}},
  \bibinfo {author} {\bibfnamefont {A.}~\bibnamefont {Rosch}},\ and\ \bibinfo
  {author} {\bibfnamefont {S.}~\bibnamefont {Diehl}},\ }\bibfield  {title}
  {\bibinfo {title} {Universal phenomenology at critical exceptional points of
  nonequilibrium {O} ({N}) models},\ }\href@noop {} {\bibfield  {journal}
  {\bibinfo  {journal} {Physical Review X}\ }\textbf {\bibinfo {volume} {14}},\
  \bibinfo {pages} {021052} (\bibinfo {year} {2024})},\ \bibinfo {note}
  {publisher: APS}\BibitemShut {NoStop}%
\bibitem [{\citenamefont
  {Golestanian}(2024)}]{golestanian_non-reciprocal_2024}%
  \BibitemOpen
  \bibfield  {author} {\bibinfo {author} {\bibfnamefont {R.}~\bibnamefont
  {Golestanian}},\ }\bibfield  {title} {\bibinfo {title} {Non-reciprocal
  active-matter: a tale of “loving hate, brawling love” across the
  scales},\ }\href@noop {} {\bibfield  {journal} {\bibinfo  {journal}
  {Europhysics News}\ }\textbf {\bibinfo {volume} {55}},\ \bibinfo {pages} {12}
  (\bibinfo {year} {2024})},\ \bibinfo {note} {publisher: EDP
  Sciences}\BibitemShut {NoStop}%
\bibitem [{\citenamefont {Avni}\ \emph {et~al.}(2024)\citenamefont {Avni},
  \citenamefont {Fruchart}, \citenamefont {Martin}, \citenamefont {Seara},\
  and\ \citenamefont {Vitelli}}]{avni_dynamical_2024}%
  \BibitemOpen
  \bibfield  {author} {\bibinfo {author} {\bibfnamefont {Y.}~\bibnamefont
  {Avni}}, \bibinfo {author} {\bibfnamefont {M.}~\bibnamefont {Fruchart}},
  \bibinfo {author} {\bibfnamefont {D.}~\bibnamefont {Martin}}, \bibinfo
  {author} {\bibfnamefont {D.}~\bibnamefont {Seara}},\ and\ \bibinfo {author}
  {\bibfnamefont {V.}~\bibnamefont {Vitelli}},\ }\bibfield  {title} {\bibinfo
  {title} {Dynamical phase transitions in the non-reciprocal {Ising} model},\
  }\href@noop {} {\bibfield  {journal} {\bibinfo  {journal} {arXiv preprint
  arXiv:2409.07481}\ } (\bibinfo {year} {2024})}\BibitemShut {NoStop}%
\bibitem [{\citenamefont {Duan}\ \emph {et~al.}(2025)\citenamefont {Duan},
  \citenamefont {Agudo-Canalejo}, \citenamefont {Golestanian},\ and\
  \citenamefont {Mahault}}]{duan_phase_2025}%
  \BibitemOpen
  \bibfield  {author} {\bibinfo {author} {\bibfnamefont {Y.}~\bibnamefont
  {Duan}}, \bibinfo {author} {\bibfnamefont {J.}~\bibnamefont
  {Agudo-Canalejo}}, \bibinfo {author} {\bibfnamefont {R.}~\bibnamefont
  {Golestanian}},\ and\ \bibinfo {author} {\bibfnamefont {B.}~\bibnamefont
  {Mahault}},\ }\bibfield  {title} {\bibinfo {title} {Phase coexistence in
  nonreciprocal quorum-sensing active matter},\ }\href@noop {} {\bibfield
  {journal} {\bibinfo  {journal} {Physical Review Research}\ }\textbf {\bibinfo
  {volume} {7}},\ \bibinfo {pages} {013234} (\bibinfo {year} {2025})},\
  \bibinfo {note} {publisher: APS}\BibitemShut {NoStop}%
\bibitem [{\citenamefont {Dadhichi}\ \emph {et~al.}(2020)\citenamefont
  {Dadhichi}, \citenamefont {Kethapelli}, \citenamefont {Chajwa}, \citenamefont
  {Ramaswamy},\ and\ \citenamefont {Maitra}}]{dadhichi_nonmutual_2020}%
  \BibitemOpen
  \bibfield  {author} {\bibinfo {author} {\bibfnamefont {L.~P.}\ \bibnamefont
  {Dadhichi}}, \bibinfo {author} {\bibfnamefont {J.}~\bibnamefont
  {Kethapelli}}, \bibinfo {author} {\bibfnamefont {R.}~\bibnamefont {Chajwa}},
  \bibinfo {author} {\bibfnamefont {S.}~\bibnamefont {Ramaswamy}},\ and\
  \bibinfo {author} {\bibfnamefont {A.}~\bibnamefont {Maitra}},\ }\bibfield
  {title} {\bibinfo {title} {Nonmutual torques and the unimportance of motility
  for long-range order in two-dimensional flocks},\ }\href@noop {} {\bibfield
  {journal} {\bibinfo  {journal} {Physical Review E}\ }\textbf {\bibinfo
  {volume} {101}},\ \bibinfo {pages} {052601} (\bibinfo {year} {2020})},\
  \bibinfo {note} {publisher: APS}\BibitemShut {NoStop}%
\bibitem [{\citenamefont {Pearce}\ \emph {et~al.}(2014)\citenamefont {Pearce},
  \citenamefont {Miller}, \citenamefont {Rowlands},\ and\ \citenamefont
  {Turner}}]{pearce_role_2014}%
  \BibitemOpen
  \bibfield  {author} {\bibinfo {author} {\bibfnamefont {D.~J.}\ \bibnamefont
  {Pearce}}, \bibinfo {author} {\bibfnamefont {A.~M.}\ \bibnamefont {Miller}},
  \bibinfo {author} {\bibfnamefont {G.}~\bibnamefont {Rowlands}},\ and\
  \bibinfo {author} {\bibfnamefont {M.~S.}\ \bibnamefont {Turner}},\ }\bibfield
   {title} {\bibinfo {title} {Role of projection in the control of bird
  flocks},\ }\href@noop {} {\bibfield  {journal} {\bibinfo  {journal}
  {Proceedings of the National Academy of Sciences}\ }\textbf {\bibinfo
  {volume} {111}},\ \bibinfo {pages} {10422} (\bibinfo {year} {2014})},\
  \bibinfo {note} {publisher: National Academy of Sciences}\BibitemShut
  {NoStop}%
\bibitem [{\citenamefont {Barberis}\ and\ \citenamefont
  {Peruani}(2016)}]{barberis_large-scale_2016}%
  \BibitemOpen
  \bibfield  {author} {\bibinfo {author} {\bibfnamefont {L.}~\bibnamefont
  {Barberis}}\ and\ \bibinfo {author} {\bibfnamefont {F.}~\bibnamefont
  {Peruani}},\ }\bibfield  {title} {\bibinfo {title} {Large-scale patterns in a
  minimal cognitive flocking model: incidental leaders, nematic patterns, and
  aggregates},\ }\href@noop {} {\bibfield  {journal} {\bibinfo  {journal}
  {Physical review letters}\ }\textbf {\bibinfo {volume} {117}},\ \bibinfo
  {pages} {248001} (\bibinfo {year} {2016})},\ \bibinfo {note} {publisher:
  APS}\BibitemShut {NoStop}%
\bibitem [{\citenamefont {Chen}\ \emph {et~al.}(2017)\citenamefont {Chen},
  \citenamefont {Patelli}, \citenamefont {Chaté}, \citenamefont {Ma},\ and\
  \citenamefont {Shi}}]{chen_fore-aft_2017}%
  \BibitemOpen
  \bibfield  {author} {\bibinfo {author} {\bibfnamefont {Q.-s.}\ \bibnamefont
  {Chen}}, \bibinfo {author} {\bibfnamefont {A.}~\bibnamefont {Patelli}},
  \bibinfo {author} {\bibfnamefont {H.}~\bibnamefont {Chaté}}, \bibinfo
  {author} {\bibfnamefont {Y.-Q.}\ \bibnamefont {Ma}},\ and\ \bibinfo {author}
  {\bibfnamefont {X.-Q.}\ \bibnamefont {Shi}},\ }\bibfield  {title} {\bibinfo
  {title} {Fore-aft asymmetric flocking},\ }\href@noop {} {\bibfield  {journal}
  {\bibinfo  {journal} {Physical Review E}\ }\textbf {\bibinfo {volume} {96}},\
  \bibinfo {pages} {020601} (\bibinfo {year} {2017})}\BibitemShut {NoStop}%
\bibitem [{\citenamefont {Romanczuk}\ \emph {et~al.}(2016)\citenamefont
  {Romanczuk}, \citenamefont {Chaté}, \citenamefont {Chen}, \citenamefont
  {Ngo},\ and\ \citenamefont {Toner}}]{romanczuk_emergent_2016}%
  \BibitemOpen
  \bibfield  {author} {\bibinfo {author} {\bibfnamefont {P.}~\bibnamefont
  {Romanczuk}}, \bibinfo {author} {\bibfnamefont {H.}~\bibnamefont {Chaté}},
  \bibinfo {author} {\bibfnamefont {L.}~\bibnamefont {Chen}}, \bibinfo {author}
  {\bibfnamefont {S.}~\bibnamefont {Ngo}},\ and\ \bibinfo {author}
  {\bibfnamefont {J.}~\bibnamefont {Toner}},\ }\bibfield  {title} {\bibinfo
  {title} {Emergent smectic order in simple active particle models},\
  }\href@noop {} {\bibfield  {journal} {\bibinfo  {journal} {New Journal of
  Physics}\ }\textbf {\bibinfo {volume} {18}},\ \bibinfo {pages} {063015}
  (\bibinfo {year} {2016})},\ \bibinfo {note} {publisher: IOP
  Publishing}\BibitemShut {NoStop}%
\bibitem [{\citenamefont {Fruchart}\ \emph {et~al.}(2023)\citenamefont
  {Fruchart}, \citenamefont {Scheibner},\ and\ \citenamefont
  {Vitelli}}]{fruchart_odd_2023}%
  \BibitemOpen
  \bibfield  {author} {\bibinfo {author} {\bibfnamefont {M.}~\bibnamefont
  {Fruchart}}, \bibinfo {author} {\bibfnamefont {C.}~\bibnamefont
  {Scheibner}},\ and\ \bibinfo {author} {\bibfnamefont {V.}~\bibnamefont
  {Vitelli}},\ }\bibfield  {title} {\bibinfo {title} {Odd viscosity and odd
  elasticity},\ }\href@noop {} {\bibfield  {journal} {\bibinfo  {journal}
  {Annual Review of Condensed Matter Physics}\ }\textbf {\bibinfo {volume}
  {14}},\ \bibinfo {pages} {471} (\bibinfo {year} {2023})},\ \bibinfo {note}
  {publisher: Annual Reviews}\BibitemShut {NoStop}%
\bibitem [{\citenamefont {Loos}\ \emph {et~al.}(2023)\citenamefont {Loos},
  \citenamefont {Klapp},\ and\ \citenamefont
  {Martynec}}]{loos_long-range_2023}%
  \BibitemOpen
  \bibfield  {author} {\bibinfo {author} {\bibfnamefont {S.~A.}\ \bibnamefont
  {Loos}}, \bibinfo {author} {\bibfnamefont {S.~H.}\ \bibnamefont {Klapp}},\
  and\ \bibinfo {author} {\bibfnamefont {T.}~\bibnamefont {Martynec}},\
  }\bibfield  {title} {\bibinfo {title} {Long-range order and directional
  defect propagation in the nonreciprocal {XY} model with vision cone
  interactions},\ }\href@noop {} {\bibfield  {journal} {\bibinfo  {journal}
  {Physical review letters}\ }\textbf {\bibinfo {volume} {130}},\ \bibinfo
  {pages} {198301} (\bibinfo {year} {2023})},\ \bibinfo {note} {publisher:
  APS}\BibitemShut {NoStop}%
\bibitem [{\citenamefont {Bandini}\ \emph {et~al.}(2024)\citenamefont
  {Bandini}, \citenamefont {Venturelli}, \citenamefont {Loos}, \citenamefont
  {Jelic},\ and\ \citenamefont {Gambassi}}]{bandini_xy_2024}%
  \BibitemOpen
  \bibfield  {author} {\bibinfo {author} {\bibfnamefont {G.}~\bibnamefont
  {Bandini}}, \bibinfo {author} {\bibfnamefont {D.}~\bibnamefont {Venturelli}},
  \bibinfo {author} {\bibfnamefont {S.~A.}\ \bibnamefont {Loos}}, \bibinfo
  {author} {\bibfnamefont {A.}~\bibnamefont {Jelic}},\ and\ \bibinfo {author}
  {\bibfnamefont {A.}~\bibnamefont {Gambassi}},\ }\bibfield  {title} {\bibinfo
  {title} {The {XY} model with vision cone: non-reciprocal vs. reciprocal
  interactions},\ }\href@noop {} {\bibfield  {journal} {\bibinfo  {journal}
  {arXiv preprint arXiv:2412.19297}\ } (\bibinfo {year} {2024})}\BibitemShut
  {NoStop}%
\bibitem [{\citenamefont {Liu}\ \emph {et~al.}(2025)\citenamefont {Liu},
  \citenamefont {Zheng}, \citenamefont {Nian},\ and\ \citenamefont
  {Xiong}}]{liu_dynamic_2025}%
  \BibitemOpen
  \bibfield  {author} {\bibinfo {author} {\bibfnamefont {Z.-Y.}\ \bibnamefont
  {Liu}}, \bibinfo {author} {\bibfnamefont {B.}~\bibnamefont {Zheng}}, \bibinfo
  {author} {\bibfnamefont {L.-L.}\ \bibnamefont {Nian}},\ and\ \bibinfo
  {author} {\bibfnamefont {L.}~\bibnamefont {Xiong}},\ }\bibfield  {title}
  {\bibinfo {title} {Dynamic approach to the two-dimensional nonreciprocal {XY}
  model with vision cone interactions},\ }\href@noop {} {\bibfield  {journal}
  {\bibinfo  {journal} {Physical Review E}\ }\textbf {\bibinfo {volume}
  {111}},\ \bibinfo {pages} {014131} (\bibinfo {year} {2025})},\ \bibinfo
  {note} {publisher: APS}\BibitemShut {NoStop}%
\bibitem [{\citenamefont {Rouzaire}\ \emph {et~al.}(2024)\citenamefont
  {Rouzaire}, \citenamefont {Pearce}, \citenamefont {Pagonabarraga},\ and\
  \citenamefont {Levis}}]{rouzaire_non-reciprocal_2024}%
  \BibitemOpen
  \bibfield  {author} {\bibinfo {author} {\bibfnamefont {Y.}~\bibnamefont
  {Rouzaire}}, \bibinfo {author} {\bibfnamefont {D.~J.}\ \bibnamefont
  {Pearce}}, \bibinfo {author} {\bibfnamefont {I.}~\bibnamefont
  {Pagonabarraga}},\ and\ \bibinfo {author} {\bibfnamefont {D.}~\bibnamefont
  {Levis}},\ }\bibfield  {title} {\bibinfo {title} {Non-{Reciprocal}
  {Interactions} {Reshape} {Topological} {Defect} {Annihilation}},\ }\href@noop
  {} {\bibfield  {journal} {\bibinfo  {journal} {arXiv preprint
  arXiv:2401.12637}\ } (\bibinfo {year} {2024})}\BibitemShut {NoStop}%
\bibitem [{\citenamefont {Toner}(2012)}]{toner_birth_2012}%
  \BibitemOpen
  \bibfield  {author} {\bibinfo {author} {\bibfnamefont {J.}~\bibnamefont
  {Toner}},\ }\bibfield  {title} {\bibinfo {title} {Birth, death, and flight:
  {A} theory of malthusian flocks},\ }\href@noop {} {\bibfield  {journal}
  {\bibinfo  {journal} {Physical review letters}\ }\textbf {\bibinfo {volume}
  {108}},\ \bibinfo {pages} {088102} (\bibinfo {year} {2012})},\ \bibinfo
  {note} {publisher: APS}\BibitemShut {NoStop}%
\bibitem [{\citenamefont {Besse}\ \emph {et~al.}(2022)\citenamefont {Besse},
  \citenamefont {Chaté},\ and\ \citenamefont
  {Solon}}]{besse_metastability_2022}%
  \BibitemOpen
  \bibfield  {author} {\bibinfo {author} {\bibfnamefont {M.}~\bibnamefont
  {Besse}}, \bibinfo {author} {\bibfnamefont {H.}~\bibnamefont {Chaté}},\ and\
  \bibinfo {author} {\bibfnamefont {A.}~\bibnamefont {Solon}},\ }\bibfield
  {title} {\bibinfo {title} {Metastability of {Constant}-{Density} {Flocks}},\
  }\href@noop {} {\bibfield  {journal} {\bibinfo  {journal} {Physical Review
  Letters}\ }\textbf {\bibinfo {volume} {129}},\ \bibinfo {pages} {268003}
  (\bibinfo {year} {2022})},\ \bibinfo {note} {publisher: APS}\BibitemShut
  {NoStop}%
\bibitem [{\citenamefont {Chaté}\ and\ \citenamefont
  {Solon}(2024)}]{chate_dynamic_2024}%
  \BibitemOpen
  \bibfield  {author} {\bibinfo {author} {\bibfnamefont {H.}~\bibnamefont
  {Chaté}}\ and\ \bibinfo {author} {\bibfnamefont {A.}~\bibnamefont {Solon}},\
  }\bibfield  {title} {\bibinfo {title} {Dynamic scaling of two-dimensional
  polar flocks},\ }\href@noop {} {\bibfield  {journal} {\bibinfo  {journal}
  {Physical Review Letters}\ }\textbf {\bibinfo {volume} {132}},\ \bibinfo
  {pages} {268302} (\bibinfo {year} {2024})},\ \bibinfo {note} {publisher:
  APS}\BibitemShut {NoStop}%
\bibitem [{\citenamefont {Chatterjee}\ \emph {et~al.}(2022)\citenamefont
  {Chatterjee}, \citenamefont {Mangeat},\ and\ \citenamefont
  {Rieger}}]{chatterjee_polar_2022}%
  \BibitemOpen
  \bibfield  {author} {\bibinfo {author} {\bibfnamefont {S.}~\bibnamefont
  {Chatterjee}}, \bibinfo {author} {\bibfnamefont {M.}~\bibnamefont
  {Mangeat}},\ and\ \bibinfo {author} {\bibfnamefont {H.}~\bibnamefont
  {Rieger}},\ }\bibfield  {title} {\bibinfo {title} {Polar flocks with
  discretized directions: {The} active clock model approaching the {Vicsek}
  model},\ }\href@noop {} {\bibfield  {journal} {\bibinfo  {journal}
  {Europhysics Letters}\ }\textbf {\bibinfo {volume} {138}},\ \bibinfo {pages}
  {41001} (\bibinfo {year} {2022})},\ \bibinfo {note} {publisher: IOP
  Publishing}\BibitemShut {NoStop}%
\bibitem [{\citenamefont {Solon}\ \emph {et~al.}(2022)\citenamefont {Solon},
  \citenamefont {Chaté}, \citenamefont {Toner},\ and\ \citenamefont
  {Tailleur}}]{solon_susceptibility_2022}%
  \BibitemOpen
  \bibfield  {author} {\bibinfo {author} {\bibfnamefont {A.}~\bibnamefont
  {Solon}}, \bibinfo {author} {\bibfnamefont {H.}~\bibnamefont {Chaté}},
  \bibinfo {author} {\bibfnamefont {J.}~\bibnamefont {Toner}},\ and\ \bibinfo
  {author} {\bibfnamefont {J.}~\bibnamefont {Tailleur}},\ }\bibfield  {title}
  {\bibinfo {title} {Susceptibility of polar flocks to spatial anisotropy},\
  }\href@noop {} {\bibfield  {journal} {\bibinfo  {journal} {Physical Review
  Letters}\ }\textbf {\bibinfo {volume} {128}},\ \bibinfo {pages} {208004}
  (\bibinfo {year} {2022})},\ \bibinfo {note} {publisher: APS}\BibitemShut
  {NoStop}%
\bibitem [{Note1()}]{Note1}%
  \BibitemOpen
  \bibinfo {note} {The equilibrium XY model at temperature $T$ is recovered
  when $J$ is constant.}\BibitemShut {Stop}%
\bibitem [{\citenamefont {Peshkov}\ \emph {et~al.}(2014)\citenamefont
  {Peshkov}, \citenamefont {Bertin}, \citenamefont {Ginelli},\ and\
  \citenamefont {Chaté}}]{peshkov_boltzmann-ginzburg-landau_2014}%
  \BibitemOpen
  \bibfield  {author} {\bibinfo {author} {\bibfnamefont {A.}~\bibnamefont
  {Peshkov}}, \bibinfo {author} {\bibfnamefont {E.}~\bibnamefont {Bertin}},
  \bibinfo {author} {\bibfnamefont {F.}~\bibnamefont {Ginelli}},\ and\ \bibinfo
  {author} {\bibfnamefont {H.}~\bibnamefont {Chaté}},\ }\bibfield  {title}
  {\bibinfo {title} {Boltzmann-{Ginzburg}-{Landau} approach for continuous
  descriptions of generic {Vicsek}-like models},\ }\href@noop {} {\bibfield
  {journal} {\bibinfo  {journal} {The European Physical Journal Special
  Topics}\ }\textbf {\bibinfo {volume} {223}},\ \bibinfo {pages} {1315}
  (\bibinfo {year} {2014})}\BibitemShut {NoStop}%
\bibitem [{Note2()}]{Note2}%
  \BibitemOpen
  \bibinfo {note} {Above we omitted a term $\sim J_2 (\partial _x^2-\partial
  _y^2)(-m_x,m_y)$ which has negligible effects and vanishes for the coupling
  function without nematic harmonics.}\BibitemShut {Stop}%
\bibitem [{\citenamefont {Toner}\ and\ \citenamefont
  {Tu}(1995)}]{toner_long-range_1995}%
  \BibitemOpen
  \bibfield  {author} {\bibinfo {author} {\bibfnamefont {J.}~\bibnamefont
  {Toner}}\ and\ \bibinfo {author} {\bibfnamefont {Y.}~\bibnamefont {Tu}},\
  }\bibfield  {title} {\bibinfo {title} {Long-range order in a two-dimensional
  dynamical {XY} model: how birds fly together},\ }\href@noop {} {\bibfield
  {journal} {\bibinfo  {journal} {Physical Review Letters}\ }\textbf {\bibinfo
  {volume} {75}},\ \bibinfo {pages} {4326} (\bibinfo {year}
  {1995})}\BibitemShut {NoStop}%
\bibitem [{SUP()}]{SUPP}%
  \BibitemOpen
  \href@noop {} {}\bibinfo {note} {See Supplemental Material
  [url].}\BibitemShut {Stop}%
\bibitem [{Note3()}]{Note3}%
  \BibitemOpen
  \bibinfo {note} {Defined as $\xi =\DOTSI \intop \ilimits@ _0^\pi
  S(q)dq/\DOTSI \intop \ilimits@ _0^\pi qS(q)dq$ with the structure factor
  $S(q)=\langle \protect \mathbf {m}(\protect \mathbf {q})\protect \mathbf
  {m}(-q)\rangle $ averaged over time and orientations of $\protect \mathbf
  {q}$.}\BibitemShut {Stop}%
\bibitem [{\citenamefont {Martin}\ \emph {et~al.}(2021)\citenamefont {Martin},
  \citenamefont {Chaté}, \citenamefont {Nardini}, \citenamefont {Solon},
  \citenamefont {Tailleur},\ and\ \citenamefont
  {Van~Wijland}}]{martin_fluctuation-induced_2021}%
  \BibitemOpen
  \bibfield  {author} {\bibinfo {author} {\bibfnamefont {D.}~\bibnamefont
  {Martin}}, \bibinfo {author} {\bibfnamefont {H.}~\bibnamefont {Chaté}},
  \bibinfo {author} {\bibfnamefont {C.}~\bibnamefont {Nardini}}, \bibinfo
  {author} {\bibfnamefont {A.}~\bibnamefont {Solon}}, \bibinfo {author}
  {\bibfnamefont {J.}~\bibnamefont {Tailleur}},\ and\ \bibinfo {author}
  {\bibfnamefont {F.}~\bibnamefont {Van~Wijland}},\ }\bibfield  {title}
  {\bibinfo {title} {Fluctuation-induced phase separation in metric and
  topological models of collective motion},\ }\href@noop {} {\bibfield
  {journal} {\bibinfo  {journal} {Physical Review Letters}\ }\textbf {\bibinfo
  {volume} {126}},\ \bibinfo {pages} {148001} (\bibinfo {year} {2021})},\
  \bibinfo {note} {publisher: APS}\BibitemShut {NoStop}%
\bibitem [{\citenamefont {Martin}\ \emph {et~al.}(2024)\citenamefont {Martin},
  \citenamefont {Spera}, \citenamefont {Chaté}, \citenamefont {Duclut},
  \citenamefont {Nardini}, \citenamefont {Tailleur},\ and\ \citenamefont {van
  Wijland}}]{martin_fluctuation-induced_2024}%
  \BibitemOpen
  \bibfield  {author} {\bibinfo {author} {\bibfnamefont {D.}~\bibnamefont
  {Martin}}, \bibinfo {author} {\bibfnamefont {G.}~\bibnamefont {Spera}},
  \bibinfo {author} {\bibfnamefont {H.}~\bibnamefont {Chaté}}, \bibinfo
  {author} {\bibfnamefont {C.}~\bibnamefont {Duclut}}, \bibinfo {author}
  {\bibfnamefont {C.}~\bibnamefont {Nardini}}, \bibinfo {author} {\bibfnamefont
  {J.}~\bibnamefont {Tailleur}},\ and\ \bibinfo {author} {\bibfnamefont
  {F.}~\bibnamefont {van Wijland}},\ }\bibfield  {title} {\bibinfo {title}
  {Fluctuation-induced first order transition to collective motion},\
  }\href@noop {} {\bibfield  {journal} {\bibinfo  {journal} {Journal of
  Statistical Mechanics: Theory and Experiment}\ }\textbf {\bibinfo {volume}
  {2024}},\ \bibinfo {pages} {084003} (\bibinfo {year} {2024})},\ \bibinfo
  {note} {publisher: IOP Publishing}\BibitemShut {NoStop}%
\bibitem [{Note4()}]{Note4}%
  \BibitemOpen
  \bibinfo {note} {Interestingly, this depends on the type of dynamics: For
  Monte-Carlo dynamics in which the angle can make arbitrary jumps at each time
  step, the pinning directions are opposite~\cite
  {loos_long-range_2023,bandini_xy_2024} compared to our Langevin dynamics
  Eq.~(\ref {eq:model}).}\BibitemShut {Stop}%
\bibitem [{\citenamefont {José}\ \emph {et~al.}(1977)\citenamefont {José},
  \citenamefont {Kadanoff}, \citenamefont {Kirkpatrick},\ and\ \citenamefont
  {Nelson}}]{jose_renormalization_1977}%
  \BibitemOpen
  \bibfield  {author} {\bibinfo {author} {\bibfnamefont {J.~V.}\ \bibnamefont
  {José}}, \bibinfo {author} {\bibfnamefont {L.~P.}\ \bibnamefont {Kadanoff}},
  \bibinfo {author} {\bibfnamefont {S.}~\bibnamefont {Kirkpatrick}},\ and\
  \bibinfo {author} {\bibfnamefont {D.~R.}\ \bibnamefont {Nelson}},\ }\bibfield
   {title} {\bibinfo {title} {Renormalization, vortices, and symmetry-breaking
  perturbations in the two-dimensional planar model},\ }\href@noop {}
  {\bibfield  {journal} {\bibinfo  {journal} {Physical Review B}\ }\textbf
  {\bibinfo {volume} {16}},\ \bibinfo {pages} {1217} (\bibinfo {year}
  {1977})},\ \bibinfo {note} {publisher: APS}\BibitemShut {NoStop}%
\bibitem [{\citenamefont {Lapilli}\ \emph {et~al.}(2006)\citenamefont
  {Lapilli}, \citenamefont {Pfeifer},\ and\ \citenamefont
  {Wexler}}]{lapilli_universality_2006}%
  \BibitemOpen
  \bibfield  {author} {\bibinfo {author} {\bibfnamefont {C.~M.}\ \bibnamefont
  {Lapilli}}, \bibinfo {author} {\bibfnamefont {P.}~\bibnamefont {Pfeifer}},\
  and\ \bibinfo {author} {\bibfnamefont {C.}~\bibnamefont {Wexler}},\
  }\bibfield  {title} {\bibinfo {title} {Universality away from critical points
  in two-dimensional phase transitions},\ }\href@noop {} {\bibfield  {journal}
  {\bibinfo  {journal} {Physical review letters}\ }\textbf {\bibinfo {volume}
  {96}},\ \bibinfo {pages} {140603} (\bibinfo {year} {2006})},\ \bibinfo {note}
  {publisher: APS}\BibitemShut {NoStop}%
\bibitem [{\citenamefont {Popli}\ \emph {et~al.}(2025)\citenamefont {Popli},
  \citenamefont {Maitra},\ and\ \citenamefont {Ramaswamy}}]{popli_dont_2025}%
  \BibitemOpen
  \bibfield  {author} {\bibinfo {author} {\bibfnamefont {P.}~\bibnamefont
  {Popli}}, \bibinfo {author} {\bibfnamefont {A.}~\bibnamefont {Maitra}},\ and\
  \bibinfo {author} {\bibfnamefont {S.}~\bibnamefont {Ramaswamy}},\ }\bibfield
  {title} {\bibinfo {title} {Don't look back: {Ordering} and defect cloaking in
  non-reciprocal lattice {XY} models},\ }\href@noop {} {\bibfield  {journal}
  {\bibinfo  {journal} {arXiv preprint arXiv:2503.06480}\ } (\bibinfo {year}
  {2025})}\BibitemShut {NoStop}%
\end{thebibliography}%

\end{document}